\documentclass[aps, prc, floatfix, nofootinbib, superscriptaddress, twocolumn]{revtex4-1}

\usepackage{latexsym}
\usepackage{amsmath}
\usepackage{amssymb}
\usepackage{amsfonts}

\usepackage[mathscr,scaled=1.15]{urwchancal}
\DeclareFontFamily{OT1}{pzc}{}
\DeclareFontShape{OT1}{pzc}{m}{it}%
{<-> s * [1.15] pzcmi7t}{}
\DeclareMathAlphabet{\mathpzc}{OT1}{pzc}{m}{it}

\usepackage{color}

\usepackage{supertabular}
\usepackage{placeins}
\usepackage{epsfig}
\usepackage{graphicx}

\definecolor{purple}{rgb}{0.5,0,0.5}
\definecolor{blue}{rgb}{0.0,0,0.9}
\definecolor{prdblue}{rgb}{0.133,0.118,0.498}
\usepackage[colorlinks=true, pdfstartview=FitV, linkcolor=prdblue, citecolor= prdblue, urlcolor=prdblue]{hyperref}

\begin{document}

\title{Light-meson masses in an unquenched quark model}

\author{Xiaoyun Chen}
\email[]{xychen@jit.edu.cn}
\affiliation{Department of Basic Courses, Jinling Institute of Technology, Nanjing 211169, P. R. China}

\author{Jialun Ping}
\email[]{jlping@njnu.edu.cn}
\affiliation{Department of Physics, Nanjing Normal University, Nanjing 210023, P. R. China}

\author{Craig D. Roberts}
\email[]{cdroberts@anl.gov}
\affiliation{Physics Division, Argonne National Laboratory, Argonne, Illinois
60439, USA}

\author{Jorge Segovia}
\email[]{jsegovia@ifae.es}
\affiliation{
Institut de F\'isica d'Altes Energies (IFAE) and Barcelona Institute of Science and Technology (BIST),
Universitat Aut\`onoma de Barcelona, E-08193 Bellaterra (Barcelona), Spain
}

\date{4 December 2017}

\begin{abstract}
We perform a coupled-channels calculation of the masses of light mesons with the quantum numbers $IJ^{P=-}$, $(I,J)=0,1$, by including $q\bar{q}$ and $(q\bar{q})^2$ components in a nonrelativistic chiral quark model.  The coupling between two- and four-quark configurations is realized through a $^3P_0$ quark-pair creation model.  With the usual form of this operator, the mass shifts are large and negative, an outcome which raises serious issues of validity for the quenched quark model.  Herein, therefore, we introduce some improvements of the $^3P_0$ operator in order to reduce the size of the mass shifts.  By introducing two simple factors, physically well motivated, the coupling between $q\bar{q}$ and $(q\bar{q})^2$ components is weakened, producing mass shifts that are around 10-20\% of hadron bare masses.  


\end{abstract}


\maketitle


\section{Introduction} \label{introduction}

In the conventional quenched quark model, a meson is described as a quark-antiquark bound state.  This picture was successfully applied to heavy quarkonia, such as bottomonium and charmonium \cite{Eichten:1978tg, Eichten:1979ms, Gupta:1982kp, Gupta:1983we, Gupta:1984jb, Gupta:1984um, Kwong:1987ak, Kwong:1988ae, Barnes:1996ff, Ebert:2002pp, Radford:2007vd, Eichten:2007qx, Segovia:2008zz, Danilkin:2009hr, Ferretti:2013vua, Segovia:2013wma, Godfrey:2015dia, Segovia:2016xqb}, and also, to some extent, light mesons \cite{Godfrey:1985xj, Vijande:2004he, Segovia:2008zza}.  However, since the discovery of the $X(3872)$ \cite{Choi:2003ue}, a large number of so-called $XYZ$ particles have been found \cite{Brambilla:2010cs}.  Some of them, especially the charged states associated with heavy quarkonium \cite{Belle:2011aa, Ablikim:2013mio}, are clear indications that there exist mesons beyond those which can be built simply from a valence-quark and -antiquark. 

The effects of hadron loops on hadron properties have been studied extensively in the framework of the coupled-channels method \cite{Tornqvist:1979hx, Ono:1983rd, Tornqvist:1984xz, Ping:2012zz, Ortega:2016mms, Ortega:2016pgg, Ortega:2016hde, Ortega:2017qmg} within the ``unquenched'' quark model.  Amongst other things, the loops can add continuum components to a bare (undressed) quark-model state, shifting its mass, producing a width, and thereby creating a ``physical'' hadron that is a considerably more complex object.  For example, in Ref.\,\cite{Geiger:1989yc}, using a $^3P_0$ model to generate the couplings \cite{Micu:1968mk, LeYaouanc:1972vsx, LeYaouanc:1973ldf}, virtual $q\bar q$ pairs were found to induce very large mass shifts; and similarly, in Ref.\,\cite{Barnes:2007xu}, large shifts ($\sim 500\,$MeV) were also induced by inclusion of all six $D$, $D^\ast$, $D_s$ and $D^\ast_s$ pair channels in the analysis of $J^{PC}=1^{--}$ $c\bar{c}$ states (the $J/\psi$ family).

The most widely discussed new state in the charmonium sector is the $X(3872)$.  As this state lies at the $D\bar{D}^\ast$ threshold, it has been suggested that the $X(3872)$ is a purely molecular $D\bar{D}^\ast$ system.  However, some recent studies indicate that the $X(3872)$  might more accurately be described as a mixture of a bare $c\bar{c}$ state and a $D\bar{D}^\ast$ molecule.  For instance, in Ref.\,\cite{Ortega:2009hj} a coupled-channels analysis of the $1^{++}$ $c\bar{c}$ sector, using a $^3P_0$ pair creation model to connect $q\bar{q}$ and $DD^\ast$ molecular configurations, revealed that the $X(3872)$ can emerge in chiral quark models as a dynamically generated mixture of $DD^*$ molecule and the $\chi_{c_1}(2P)$, where the $c\bar{c}$ component represents less than 10\% of the compound system; and Ref.\,\cite{Li:2009ad} found the $X(3872)$ system to be a $\chi_{c_1}(2P)$-dominated charmonium state in two different frameworks: a coupled-channels model and a screening-potential model.

In addition to the observed charmonium and charmonium-related states, many bottomonium states have also been reported, \emph{e.g}.\ $\eta_b(1S)$ \cite{Aubert:2008ba}, $\Upsilon(^3D_J)$ \cite{delAmoSanchez:2010kz},
$h_b(1P)$ \cite{Lees:2011zp}, $h_b(2P)$ \cite{Adachi:2011ji}, etc.  Hadron loop effects have been investigated in this connection \cite{Liu:2011yp}, too, with a $ ^3P_0$ model used to describe the constituent $b\bar b$ system's coupling to the two-meson $B \bar B$ continuum, where $B(\bar B)$ denotes $B(\bar{B})$, $B_s(\bar{B}_s)$, $B^\ast(\bar{B}^\ast)$ or $B^\ast_s(\bar{B}^\ast_s)$.  In this case mass-shifts of around 100\,MeV are found, so that the effects are smaller than in the charmonium sector.

Evidently, the $^3 P_0$ pair-creation model is that most widely used in quark model explorations of coupled-channel effects in the heavy-quark sector.  It has also been used to study the strong decays of light-quark mesons and baryons \cite{Geiger:1991qe, Blundell:1995ev}.  In standard versions of this model \cite{LeYaouanc:1972vsx, LeYaouanc:1973ldf}, the quark-antiquark pair is created with ``vacuum quantum numbers'', \emph{viz}.\ $J^{PC}=0^{++}$, and the probability of creation is assumed to be independent of the pair's position and energy.  With these assumptions, coupled-channels analyses of hadron loop effects normally produce alarmingly large mass shifts.  Were such an outcome unavoidable, then it would seriously undermine the validity of the quenched quark model.  Such a conclusion, however, is contradicted by the wide-ranging phenomenological success of the quenched quark model.  We are thus led to pose two questions: Is the $ ^3P_0$ model a valid foundation for the study of coupled-channels effects; and are large mass-shifts physically reasonable?

In order to address these issues herein, we compute the spectrum of $IJ^{P=-}$, $(I,J)=0,1$ light-mesons, incorporating hadron loops in a chiral quark model and solving the quantum mechanics problem using the Gaussian expansion method (GEM) \cite{Hiyama:2003cu}.  We use a $ ^3P_0$ model to describe pair creation, but deliberately explore the impact of physically motivated modifications of the associated operator.
Our Hamiltonian and method for solving for the coupled $q\bar q$, $(q\bar{q})^2$ systems are detailed in Sec.\,\ref{GEM and chiral quark model}; implementation of the $^3P_0$ model is explained in Sec.\,\ref{sec3P0}; Sec.\,\ref{Numerical Results} is devoted to a discussion of the results; and Sec.\,\ref{epilogue} is a summary.


\section{Chiral quark model and GEM}
\label{GEM and chiral quark model}
%
%
In the chiral quark model \cite{Valcarce:2005em}, the meson spectrum is obtained by solving a Schr\"{o}dinger equation:
\begin{equation}
\label{Hamiltonian1}
H \Psi_{M_I M_J}^{IJ} (1,2) =E^{IJ} \Psi_{M_I M_J}^{IJ} (1,2)\,,
\end{equation}
where $1$, $2$ are particle labels.  The wave function of a meson with quantum numbers $IJ^{PC}$ can be written:
\begin{align}
\nonumber
& \Psi_{M_I M_J}^{IJ}(1,2) \\
& =\sum_{\alpha}C_{\alpha} \left[ \psi_{l}(\mathbf{r})\chi_{s}(1,2)\right]^{J}_{M_J} \omega^c(1,2)\phi^I_{M_I}(1,2),
\label{PsiIJM}
\end{align}
where $\alpha$ denotes the intermediate quantum numbers, $l,s$ and possible flavor indices (for $I=0$ states, these indices take the values $u\bar{u}, d\bar{d}$ and $s\bar{s}$); the bracket ``[\;]'' indicates angular momentum coupling; and $\chi_{s}(1,2)$, $\omega^c(1,2)$, $\phi^I(1,2)$ are spin, color and flavor wave functions, respectively (with specific meson isospin, $I$).

Using GEM, the spatial wave function is written as a product: radial-function$\times$spherical-harmonic, and the radial part is expanded using Gaussians:
\begin{subequations}
\label{radialpart}
\begin{align}
\psi_{lm}(\mathbf{r}) & = \sum_{n=1}^{n_{\rm max}} c_{n}\psi^G_{nlm}(\mathbf{r}),\\
\psi^G_{nlm}(\mathbf{r}) & = N_{nl}r^{l} e^{-\nu_{n}r^2}Y_{lm}(\hat{\mathbf{r}}),
\end{align}
\end{subequations}
with the Gaussian size parameters chosen according to the following geometric progression
\begin{equation}\label{gaussiansize}
\nu_{n}=\frac{1}{r^2_n}, \quad r_n=r_1a^{n-1}, \quad
a=\left(\frac{r_{n_{\rm max}}}{r_1}\right)^{\frac{1}{n_{\rm max}-1}}.
\end{equation}
This procedure enables optimization of the ranges using just a small number of Gaussians.

At this point, the wave function is expressed as follows:
\begin{align}
\nonumber
&\Psi_{M_I M_J}^{IJ}(1,2) \\
& =\sum_{n\alpha} C_{\alpha}c_n
 \left[ \psi^G_{nl}(\mathbf{r})\chi_{s}(1,2) \right]^{J}_{M_J}\omega^c(1,2)\phi^I_{M_I}(1,2).\label{Gauss1}
\end{align}
Since the Gaussians in Eq.\,\eqref{Gauss1} are not orthogonal, we employ the Rayleigh-Ritz variational principle for solving the Schr\"{o}dinger equation, which leads to a generalized eigenvalue problem
\begin{subequations}
\label{HEproblem}
\begin{align}
 \sum_{n^{\prime},\alpha^{\prime}} & (H_{n\alpha,n^{\prime}\alpha^{\prime}}^{IJ}
-E^{IJ} N_{n\alpha,n^{\prime}\alpha^{\prime}}^{IJ}) C_{n^{\prime}\alpha^{\prime}}^{IJ} = 0, \\
 &H_{n\alpha,n^{\prime}\alpha^{\prime}}^{IJ} =
  \langle\Phi^{IJ}_{M_I M_J,n\alpha}| H | \Phi^{IJ}_{M_I M_J,n^{\prime}\alpha^{\prime}}\rangle ,\\
 &N_{n\alpha,n^{\prime}\alpha^{\prime}}^{IJ}=
  \langle\Phi^{IJ}_{M_I M_J,n\alpha}|1| \Phi^{IJ}_{M_I M_J,n^{\prime}\alpha^{\prime}}\rangle,
\end{align}
\end{subequations}
with
$\Phi^{IJ}_{M_I M_J,n\alpha} = [\psi^G_{nl}(\mathbf{r})\chi_{s}(1,2) ]^{J}_{M_J} \omega^c(1,2)\phi^I_{M_I}(1,2)$,
$C_{n\alpha}^{IJ} = C_{\alpha}c_n =: {\mathpzc C}^{\alpha}_n$.

The mass of the $(q\bar{q})^2$  state is also obtained by solving a Schr\"{o}dinger equation:
\begin{equation}
    H \, \Psi^{4\,IJ}_{M_IM_J}=E^{IJ} \Psi^{4\,IJ}_{M_IM_J},
\end{equation}
where $\Psi^{4\,IJ}_{M_IM_J}$ is the wave function of the four-quark state, which can be constructed as follows.
First, one writes the wave functions of two clusters, here taking a meson-meson configuration as an example:
\begin{subequations}
\begin{align}
\nonumber
&    \Psi^{I_1J_1}_{M_{I_1}M_{J_1}}(1,2)=\sum_{\alpha_1 n_1} {\mathpzc C}^{\alpha_1}_{n_1} \\
    & \times  \left[ \psi^G_{n_1 l_1}(\mathbf{r}_{12})\chi_{s_1}(1,2)\right]^{J_1}_{M_{J_1}}
 \omega^{c_1}(1,2)\phi^{I_1}_{M_{I_1}}(1,2),   \\
&    \Psi^{I_2J_2}_{M_{I_2}M_{J_2}}(3,4)=\sum_{\alpha_2 n_2} {\mathpzc C}^{\alpha_2}_{n_2} \nonumber \\
    & \times \left[ \psi^G_{n_2 l_2}(\mathbf{r}_{34})\chi_{s_2}(3,4)\right]^{J_2}_{M_{J_2}}
    \omega^{c_2}(3,4)\phi^{I_2}_{M_{I_2}}(3,4),
\end{align}
\end{subequations}
where $\chi_{s}$, $\omega^c$, $\phi^{I}$ are, respectively, spin, color and flavor wave functions of the quark-antiquark cluster.  (The quarks are numbered 1, 3, and the antiquarks 2, 4.)  Then the total wave function of the four-quark state is:
\begin{align}
& \Psi^{4\,IJ}_{M_IM_J}  =  {\cal A} \sum_{L_r}\left[ \Psi^{I_1J_1}(1,2)\Psi^{I_2J_2}(3,4)
     \psi_{L_r}(\mathbf{r}_{1234})\right]^{IJ}_{M_IM_J}    \nonumber \\
\nonumber
  & =  \sum_{\alpha_1\,\alpha_2\,n_1\,n_2\,L_r}
  {\mathpzc C}^{\alpha_1}_{n_1} {\mathpzc C}^{\alpha_2}_{n_2} \bigg[ \left[\psi^G_{n_1 l_1}(\mathbf{r}_{12})\chi_{s_1}(1,2)\right]^{J_1} \nonumber \\
& \quad \times
            \left[\psi^G_{n_2 l_2}(\mathbf{r}_{34})\chi_{s_2}(3,4)\right]^{J_2}
             \psi_{L_r}(\mathbf{r}_{1234})\bigg]^{J}_{M_J} \nonumber \\
 &      \quad  \times \left[\omega^{c_1}(1,2)\omega^{c_2}(3,4)\right]^{[1]}
     \left[\phi^{I_1}(1,2)\phi^{I_2}(3,4)\right]^{I}_{M_I},
\end{align}
where $\psi_{L_r}(\mathbf{r}_{1234})$ is the two-cluster relative wave function, describing relative cluster orbital angular momentum $L_r$, which is also expanded in a series of Gaussians, and the superscript ``$[1]$'' indicates that the cluster wave functions are coupled into the color-singlet configuration.
Here, ${\cal A}$ is the antisymmetrization operator: if all quarks (antiquarks) are taken as identical particles, then
\begin{equation}
{\cal A}=\frac{1}{2}(1-P_{13}-P_{24}+P_{13}P_{24}).
\end{equation}
In this case, too, the radial part of the wave function is expanded using Gaussians, as in Eq.\,(\ref{radialpart}), with the size parameters in Eq.~(\ref{gaussiansize}).

The calculation of Hamiltonian matrix elements is complicated if any one of the relative orbital angular momenta is nonzero.  In this case, it is useful to employ the method of infinitesimally shifted Gaussians \cite{Hiyama:2003cu}, wherewith the spherical harmonics are absorbed into the Gaussians:
\begin{eqnarray}
\label{wavefunctionG}
 \psi^G_{nlm}(\mathbf{r})& =& N_{nl}r^{l}
 e^{-\nu_{n}r^2}Y_{lm}(\hat{\mathbf{r}}) \nonumber \\
 & =&N_{nl}\lim_{\epsilon \rightarrow
 0}\frac{1}{\epsilon^{l}}\sum_{k}^{k_{\rm max}}C_{lm,k} \, {\rm e}^{-\nu_{n}(\mathbf{r}-\epsilon\mathbf{D_{lm,k}})^2},
\end{eqnarray}
where, plainly, the quantities $C_{lm,k}$, $D_{lm,k}$ are fixed by the particular spherical harmonic under consideration and their values ensure the limit $\epsilon\to 0$ exists.

The Hamiltonian of the chiral quark model consists of three parts: quark rest mass, kinetic energy, and potential energy:
\begin{align}
 H & = \sum_{i=1}^4 m_i  +\frac{p_{12}^2}{2\mu_{12}}+\frac{p_{34}^2}{2\mu_{34}}
  +\frac{p_{1234}^2}{2\mu_{1234}}  \quad  \nonumber \\
  & + \sum_{i<j=1}^4 \left[ V_{ij}^{C}+V_{ij}^{G}+\sum_{\chi=\pi,K,\eta} V_{ij}^{\chi}
   +V_{ij}^{\sigma}\right].
\end{align}
The potential energy is constituted from pieces describing quark confinement, ``C''; one-gluon-exchange, ``G''; one Goldstone boson exchange, ``$\chi=\pi$, $K$, \ldots'', and $\sigma$ exchange; and the form for the four-quark states is \cite{Valcarce:2005em}:
{\allowdisplaybreaks
\begin{subequations}
\begin{align}
V_{ij}^{C}&= ( -a_c r_{ij}^2-\Delta ) \boldsymbol{\lambda}_i^c
\cdot \boldsymbol{\lambda}_j^c ,  \\
 V_{ij}^{G}&= \frac{\alpha_s}{4} \boldsymbol{\lambda}_i^c \cdot \boldsymbol{\lambda}_{j}^c
\left[\frac{1}{r_{ij}}-\frac{2\pi}{3m_im_j}\boldsymbol{\sigma}_i\cdot
\boldsymbol{\sigma}_j
  \delta(\boldsymbol{r}_{ij})\right],  \\
\delta{(\boldsymbol{r}_{ij})} & =  \frac{e^{-r_{ij}/r_0(\mu_{ij})}}{4\pi r_{ij}r_0^2(\mu_{ij})}, \\
V_{ij}^{\pi}&= \frac{g_{ch}^2}{4\pi}\frac{m_{\pi}^2}{12m_im_j}
  \frac{\Lambda_{\pi}^2}{\Lambda_{\pi}^2-m_{\pi}^2}m_\pi v_{ij}^{\pi}
  \sum_{a=1}^3 \lambda_i^a \lambda_j^a,  \\
V_{ij}^{K}&= \frac{g_{ch}^2}{4\pi}\frac{m_{K}^2}{12m_im_j}
  \frac{\Lambda_K^2}{\Lambda_K^2-m_{K}^2}m_K v_{ij}^{K}
  \sum_{a=4}^7 \lambda_i^a \lambda_j^a,   \\
\nonumber
V_{ij}^{\eta} & =
\frac{g_{ch}^2}{4\pi}\frac{m_{\eta}^2}{12m_im_j}
\frac{\Lambda_{\eta}^2}{\Lambda_{\eta}^2-m_{\eta}^2}m_{\eta}
v_{ij}^{\eta}  \\
 & \quad \times \left[\lambda_i^8 \lambda_j^8 \cos\theta_P
 - \lambda_i^0 \lambda_j^0 \sin \theta_P \right],   \\
 v_{ij}^{\chi} & =  \left[ Y(m_\chi r_{ij})-
\frac{\Lambda_{\chi}^3}{m_{\chi}^3}Y(\Lambda_{\chi} r_{ij})
\right]
\boldsymbol{\sigma}_i \cdot\boldsymbol{\sigma}_j,\\
V_{ij}^{\sigma}&= -\frac{g_{ch}^2}{4\pi}
\frac{\Lambda_{\sigma}^2}{\Lambda_{\sigma}^2-m_{\sigma}^2}m_\sigma \nonumber \\
& \quad \times \left[
 Y(m_\sigma r_{ij})-\frac{\Lambda_{\sigma}}{m_\sigma}Y(\Lambda_{\sigma} r_{ij})\right]  ,
\end{align}
\end{subequations}}
\hspace*{-0.5\parindent}%
where $Y(x)  =   e^{-x}/x$;
$\{m_i\}$ are the constituent masses of quarks and antiquarks, and $\mu_{ij}$ are their reduced masses;
\begin{equation}
\mu_{1234}=\frac{(m_1+m_2)(m_3+m_4)}{m_1+m_2+m_3+m_4};
\end{equation}
$\mathbf{p}_{ij}=(\mathbf{p}_i-\mathbf{p}_j)/2$, $\mathbf{p}_{1234}= (\mathbf{p}_{12}-\mathbf{p}_{34})/2$;
$r_0(\mu_{ij}) =s_0/\mu_{ij}$;
$\boldsymbol{\sigma}$ are the $SU(2)$ Pauli matrices;
$\boldsymbol{\lambda}$, $\boldsymbol{\lambda}^c$ are $SU(3)$ flavor, color Gell-Mann matrices, respectively;
$g^2_{ch}/4\pi$ is the chiral coupling constant, determined from the $\pi$-nucleon coupling;
and $\alpha_s$ is an effective scale-dependent running coupling \cite{Valcarce:2005em},
\begin{equation}
\alpha_s(\mu_{ij})=\frac{\alpha_0}{\ln\left[(\mu_{ij}^2+\mu_0^2)/\Lambda_0^2\right]}.
\end{equation}
All the parameters are determined by fitting the meson spectrum, from light to heavy;
and the resulting values are listed in Table~\ref{modelparameters}.

\begin{table}[!t]
\begin{center}
\caption{ \label{modelparameters}
Model parameters, determined by fitting the meson spectrum, leaving room for unquenching contributions in the case of light-quark systems.}
\begin{tabular}{llr}
\hline\noalign{\smallskip}
Quark masses   &$m_u=m_d$    &313  \\
   (MeV)       &$m_s$         &536  \\
               &$m_c$         &1728 \\
               &$m_b$         &5112 \\
\hline
Goldstone bosons   &$m_{\pi}$     &0.70  \\
   (fm$^{-1} \sim 200\,$MeV )     &$m_{\sigma}$     &3.42  \\
                   &$m_{\eta}$     &2.77  \\
                   &$m_{K}$     &2.51  \\
                   &$\Lambda_{\pi}=\Lambda_{\sigma}$     &4.2  \\
                   &$\Lambda_{\eta}=\Lambda_{K}$     &5.2  \\
                   \cline{2-3}
                   &$g_{ch}^2/(4\pi)$                &0.54  \\
                   &$\theta_p(^\circ)$                &-15 \\
\hline
Confinement        &$a_c$ (MeV fm$^{-2}$)         &101 \\
                   &$\Delta$ (MeV)     &-78.3 \\
\hline
OGE                 & $\alpha_0$        &3.67 \\
                   &$\Lambda_0({\rm fm}^{-1})$ &0.033 \\
                  &$\mu_0$(MeV)    &36.98 \\
                   &$s_0$(MeV)    &28.17 \\
\hline
\end{tabular}
\end{center}
\end{table}


\section{$^3P_0$ model}
\label{sec3P0}
The $^3P_0$ quark-pair creation model \cite{Micu:1968mk, LeYaouanc:1972vsx, LeYaouanc:1973ldf} has been widely applied to OZI rule allowed two-body strong decays of hadrons \cite{Capstick:1986bm, Roberts:1992js, Capstick:1993kb, Page:1995rh, Ackleh:1996yt, Segovia:2012cd}. The associated operator is:
\begin{align} \label{T0}
T_0 & =-3\, \gamma \sum_m\langle 1m1(-m)|00\rangle\int
d\mathbf{p}_3d\mathbf{p}_4\delta^3(\mathbf{p}_3+\mathbf{p}_4)\nonumber\\
& \quad \times{\cal{Y}}^m_1(\frac{\mathbf{p}_3-\mathbf{p}_4}{2})
\chi^{34}_{1-m}\phi^{34}_0\omega^{34}_0b^\dagger_3(\mathbf{p}_3)d^\dagger_4(\mathbf{p}_4),
\end{align}
where $\gamma$ describes the probability for creating a quark-antiquark pair with momenta $\mathbf{p}_3$ and $\mathbf{p}_4$, respectively from the $0^{++}$ vacuum, and $\omega^{34}_0$ and $\phi^{34}_{0}$  are, in turn, color- and flavor-singlet wave function components.  (The quark and the antiquark in the source meson are labeled by 1 and 2).
The matrix element for the transition $A \rightarrow B + C$ can then be written:
\begin{equation}
\label{defT42}
\langle BC|T_{42}|A\rangle=\delta^3(\mathbf{P}_A-\mathbf{P}_B-\mathbf{P}_C) \, {\cal{M}}^{M_{J_A}M_{J_B}M_{J_C}},
\end{equation}
where $\mathbf{P}_B$, $\mathbf{P}_C$ are the momenta of the $B$ and $C$ mesons that appear in the final state, with $\mathbf{P}_A = \mathbf{P}_B + \mathbf{P}_C = 0$ in the center-of-mass frame of meson $A$. ${\cal{M}}^{M_{J_A}M_{J_B}M_{J_C}}$ is the helicity amplitude for the process $A \rightarrow B + C$:
\begin{widetext}
\begin{align}
{\cal{M}}\,&{\!}^{M_{J_A}M_{J_B}M_{J_C}}(\mathbf{P})  =
\gamma\sqrt{8E_AE_BE_C}
\sum_{M_{L_i},M_{S_i},m}^{i=A,B,C}
\langle L_AM_{L_A}S_AM_{S_A}|J_AM_{J_A}\rangle
\langle L_BM_{L_B}S_BM_{S_B}|J_BM_{J_B}\rangle \rule{12em}{0ex}\nonumber \\
&
\rule{10em}{0ex} \times \langle L_CM_{L_C}S_CM_{S_C}|J_CM_{J_C}\rangle\langle 1m1(-m)|00\rangle
\langle\chi^{14}_{S_BM_{S_B}}\chi^{32}_{S_CM_{S_C}}|\chi^{12}_{S_AM_{S_A}}\chi^{34}_{1-m}\rangle
\nonumber\\
\nonumber
&
\rule{10em}{0ex} \times \bigg[ \langle \phi^{14}_B \phi^{32}_C | \phi^{12}_A\phi^{34}_0 \rangle\,
\mathcal{I}^{M_{L_A},m}_{M_{L_B},M_{L_C}}(\mathbf{P},m_1,m_2,m_3) \\
&
\rule{12em}{0ex} + (-1)^{1+S_A+S_B+S_C}\langle\phi^{32}_B\phi^{14}_C|\phi^{12}_A\phi^{34}_0\rangle
\mathcal{I}^{M_{L_A},m}_{M_{L_B},M_{L_C}}(-\mathbf{P},m_2,m_1,m_3) \bigg],
\end{align}
with the momentum space integral
\begin{align}
\nonumber \mathcal{I}^{M_{L_A},m}_{M_{L_B},M_{L_C}}& (\mathbf{P},m_1,m_2,m_3) \\
& =\int
d^3p\, \mbox{}\psi^\ast_{n_BL_BM_{L_B}}
({\scriptstyle\frac{m_3}{m_1+m_3}}\mathbf{P}+\mathbf{p})\psi^\ast_{n_CL_CM_{L_C}}
({\scriptstyle\frac{m_3}{m_2+m_3}}\mathbf{P}+\mathbf{p})
\psi_{n_AL_AM_{L_A}}
(\mathbf{P}+\mathbf{p}){\cal{Y}}^m_1(\mathbf{p}),
\label{space integral}
\end{align}
\end{widetext}
where $\mathbf{P}=\mathbf{P}_B=-\mathbf{P}_C$, $\mathbf{p}=\mathbf{p}_3$, and $m_3$ is the mass of the created quark, $q_3$.  Here, $\psi_{n L M_{L}}$ is the (Fourier transform) of the wave function in Eq.\,\eqref{PsiIJM}, which is obtained via the self-consistent solution of the Hamiltonian problem in Eq.\,\eqref{Hamiltonian1}.

The parameters in the $ ^3P_0$ model can be constrained by computing the partial decay width of one or more mesons.  For the process $A \rightarrow B + C$,
\begin{subequations}
\begin{equation}
\Gamma = \pi^2 \frac{{|\textbf{P}|}}{M_A^2}\sum_{JL}\Big
|\mathcal{M}^{J L}\Big|^2, \label{partialwidth}
\end{equation}
where nonrelativistic phase-space is assumed,
\begin{equation}
|\textbf{P}| =\frac{\sqrt{[M^2_A-(M_B+M_C)^2][M^2_A-(M_B-M_C)^2]}}{2M_A},
\end{equation}
with $M_A$, $M_B$, $M_C$ being the masses of the mesons involved, and the partial wave amplitude ${\mathcal{M}}^{JL} (A\rightarrow BC)$ is related to the helicity amplitude via \cite{Jacob:1959at}:
\begin{align}
& {\mathcal{M}}^{J L} (A\rightarrow BC)  = \frac{\sqrt{2 L+1}}{2 J_A
+1} \!\! \sum_{M_{J_B},M_{J_C}} \langle L 0 J M_{J_A}|J_A
M_{J_A}\rangle \nonumber\\
& \rule{1em}{0ex}\times \langle J_B M_{J_B}
J_C M_{J_C} | J M_{J_A} \rangle \mathcal{M}^{M_{J_A} M_{J_B}
M_{J_C}}({\textbf{P}}). \label{MJL}
\end{align}
\end{subequations}
As an example, $\gamma$ in Eq.\,\eqref{T0} is normally determined by fitting an array of hadron strong decays.  This yields $\gamma=6.95$ for $u\bar{u}$ and $d\bar{d}$ pair creation, and $\gamma=6.95/\sqrt{3}$ for $s\bar{s}$ pair creation \cite{LeYaouanc:1977gm}.  We will initially use this value, but revise it by fitting the $\rho-\pi\pi$ width when amending the $ ^3P_0$ model.


\section{Numerical Results and discussions}
\label{Numerical Results}
\subsection{Basic Framework}
In the unquenched quark model, the eigenstates of the system can also be obtained by solving the Schr\"{o}dinger
equation:
\begin{eqnarray}
H\Psi=E\Psi ,
\end{eqnarray}
where $\Psi$ is the wave function of the system, which contains two- and four-quark components:
\begin{eqnarray}
\Psi=c_1\Psi_{2q}+c_2\Psi_{4q} \,.
\end{eqnarray}

In the nonrelativistic quark model, the number of particles is conserved.  Therefore, to study coupled-channels effects herein, we proceed as follows.  The Hamiltonian is
\begin{eqnarray}
H=H_{2q}+H_{4q}+T_{42}\,,
\end{eqnarray}
where $H_{2q}$ acts only on the wave function of mesons, $\Psi_{2q}$;
$H_{4q}$ on the four-quark wave function, $\Psi_{4q}$;
and $T_{42}$ is the transition operator in the $^3P_0$ model, Eqs.~(\ref{defT42})--\eqref{space integral}, which couples the two- and four-quark  components.  The matrix elements of the Hamiltonian can then be written:
\begin{align}
\langle\Psi| & H|\Psi\rangle =
\langle c_1\Psi_{2q}+c_2\Psi_{4q}|H|c_1\Psi_{2q}+c_2\Psi_{4q}\rangle
\nonumber \\
&=c_1^2\langle\Psi_{2q}|H_{2q}|\Psi_{2q}\rangle+c_2^2\langle\Psi_{4q}|H_{4q}|\Psi_{4q}\rangle
\nonumber \\
&\quad+c_1c_2^*\langle\Psi_{4q}|T_{42}|\Psi_{2q}\rangle+c_1^*c_2\langle\Psi_{2q}|T_{42}^{\dagger}|\Psi_{4q}\rangle.
\end{align}
In this way we arrive at a block-matrix structure for the Hamiltonian and overlap:
\begin{equation}
(H)=\left[\begin{array}{cc} (H_{2q}) & (H_{24})\\
(H_{42}) & (H_{4q})
\end{array}
\right],
(N)=\left[\begin{array}{cc} (N_{2q}) & (0)\\
(0) & (N_{4q})
\end{array}
\right] \,,
\end{equation}
where
{\allowdisplaybreaks
\begin{subequations}
\begin{align}
 (H_{2q})&=\langle\Psi_{2q}|H_{2q}|\Psi_{2q}\rangle, \\
 (H_{24})&=\langle\Psi_{4q}|T_{24}|\Psi_{2q}\rangle, \\
 (H_{4q})&=\langle\Psi_{4q}|H_{4q}|\Psi_{4q}\rangle,\\
(N_{2q})&=\langle\Psi_{2q}|1|\Psi_{2q}\rangle, \\
(N_{4q})&=\langle\Psi_{4q}|1|\Psi_{4q}\rangle.
\end{align}
\end{subequations}
}
\hspace*{-0.5\parindent}The Hamiltonian diagonalization problem, Eq.\,\eqref{HEproblem}:
\begin{eqnarray}
\Big[
\begin{array}{c}
(H)-E_n(N)
\end{array}
\Big]
\Big[
\begin{array}{c} C_n
\end{array}
\Big]=0.
\label{geig}
\end{eqnarray}
is then solved to determine the eigenenergy $E_n$ and expansion coefficients $C_n$.


The operator $T_{0}$ in Eq.~(\ref{T0}) must be Fourier transformed because the two- and four-body systems are solved in coordinate space.  In doing this, we insert a convergence factor $e^{-f^2 \mathbf{p}^2}$ into the expression, writing:
\begin{eqnarray}\label{Tr}
T_f&=&-i3\gamma\sum_{m}\langle 1m1(-m)|00\rangle\int
d\mathbf{r_3}d\mathbf{r_4}(\frac{1}{2\pi})^{\frac{3}{2}}2^{-\frac{5}{2}}f^{-5} \nonumber \\
 &&rY_{1m}(\hat{\mathbf{r}})
 {\rm e}^{-\frac{\mathbf{r}^2}{4f^2}}\chi_{1-m}^{34}\omega_{0}^{34} \phi_{0}^{34}
 b_3^{\dagger}(\mathbf{r_3})d_4^{\dagger}(\mathbf{r_4}) ,
\end{eqnarray}
where $\mathbf{r} = (\mathbf{r}_3 - \mathbf{r}_4)$ and $\hat{\mathbf{r}}$ is the associated direction-vector.  With $f\to 0$ in Eq.~(\ref{Tr}), the original form of the $ ^3P_0$ quark-pair creation operator is recovered.  We remark here that the convergence factor, $\exp{(-\mathbf{r}^2/4 f^2)}$, will acquire a physical meaning below, when we develop improvements to the $^3P_0$ operator.

Upon solving Eq.\,(\ref{geig}) with the transition matrix constructed from $T_f$ in Eq.\,\eqref{Tr} and in the limit $f \to 0$,  we obtain the results for the states $\pi$, $\rho$, $\omega$ and $\eta$,  shown in Table~\ref{result1}.
Here, the $\pi$, $\rho$, $\omega$, $\eta$ bare masses were obtained in the quenched quark model, \emph{viz}.\ solved with only the $q\bar{q}$ component.
Evidently, in each case, hadron-loop effects generate alarmingly large negative mass-shifts ($\sim -2\,000\,$MeV) for all the light mesons.  Combining this observation with those obtained elsewhere, one finds the following pattern: $b\bar b$, mass-shifts $\sim -100\,$MeV \cite{Liu:2011yp}; $c\bar c$, $\sim -(300-500)\,$MeV \cite{Barnes:2007xu, Ping:2012zz}; and $n\bar n$, $\sim -2\,000\,$MeV.  %
In our view, such large shifts invalidate this straightforward approach to unquenching the quark model.  In the following, therefore, we introduce modifications to $T_{0}$ in the $^3P_0$ model in order to develop a more realistic unquenching procedure.

\begin{table}[bt]
\begin{center}
\caption{Mass shifts computed for non-strange mesons with quantum numbers $IJ^-(I=0,1;J=0,1)$ using the transition matrix constructed from $T_{f\to 0 }$ in Eq.\,\eqref{Tr}.  ($\eta$ is an isospin 0 partner to the pion; and all dimensioned quantities are listed in MeV.)}
\label{result1}
\begin{tabular}{ccccc} \hline
states$(IJ^P)$ & $\pi(10^-)$ & $\rho(11^-)$ &  $\omega(01^-)$ &  $\eta(00^-)$ \\ \hline
bare mass (Theo.) & 139.0  &  772.7   &  701.9   & 669.5 \\ \hline
$\pi\pi$ & - & $-130.1$ & -  &  -  \\
$\pi\rho$ & $-847.9$ & - & $-596.4$ & - \\
$\pi\omega$ & -  & $-182.5$ &  -  &  -  \\
$\eta\rho$ & -  & $-159.3$ &  -  &  -  \\
$\rho\rho$ &  - & $-632.2$ &  -  &  $-834.4$  \\
$\rho\omega$ & $-804.4$  &  -  &  -  &  -  \\
$\eta\omega$ &  -  &  -  &  $-175.1$  &  - \\
$\omega\omega$ &  -   &  -  &  -  & $-271.1$ \\
$K \bar{K}$ &  -  &  $-65.0$  &  $-70.7$  &  -  \\
$K\bar{K}^{\star}(\bar{K}K^{\star})$ & $-340.2$ & $-122.4$  &  $-125.3$  & $-214.0$ \\
$K^{\star}\bar{K}^{\star}$ & $-680.0$  &  $-450.9$  &  $-506.2$  &  $-421.8$  \\ \hline
Total mass shift & ~$-2672.5$~  & ~$-1742.4$~  & ~$-1473.7$~  & ~$-1741.3$~  \\
 \hline
\end{tabular}
\end{center}
\end{table}

\subsection{Improvements}
\subsubsection{Improvement One}
In unquenching bare quark-model composites, the role played by two-meson intermediate states should diminish as their momentum, $\mathbf{p}$,  increases.  (Such a feature is seen in quantum field theory treatments of these effects \cite{Pichowsky:1999mu}.)  Hence, our first modification of the $^3P_0$ model is to reinterpret the convergence factor introduced above as a physically required feature of a realistic unquenching procedure.  Namely, we redefine $T_0\to T_1$,
\begin{align}
&T_1=-3 \,
\gamma\sum_m\langle 1m1(-m)|00\rangle\int
d\mathbf{p}_3d\mathbf{p}_4\delta^3(\mathbf{p}_3+\mathbf{p}_4)
\nonumber \\
& \quad
\times {\cal{Y}}^m_1( \hat{\mathbf{p}}) \, {\rm e}^{-f^2 \mathbf{p}^2}
\chi^{34}_{1-m}\phi^{34}_0\omega^{34}_0b^\dagger_3(\mathbf{p}_3)d^\dagger_4(\mathbf{p}_4),
\label{Tp}
\end{align}
where $\mathbf{p} = (\mathbf{p}_3-\mathbf{p}_4)/2$ is the relative momentum of the quark pair.  The coordinate-space form of Eq.\,(\ref{Tp}) is just Eq.\,(\ref{Tr}); and now, $f$ is a parameter, upon which depend our mass-shift predictions.  Their sensitivity is exhibited in Table~\ref{result2}: when $f$ is assigned a value commensurate with natural hadronic scales, $f\in [0.3,0.7]$\,fm, unquenching effects are modest; and they vanish as $f$ increases.

\begin{table}[bt]
\begin{center}
\caption{$\pi\rho$ contribution to $\pi$ mass, computed with the modified
transition operator in Eq.~(\ref{Tp}). (Unit: MeV)} \label{result2}
\begin{tabular}{cccccccc} \hline
$f$ (fm)  &  0.001 & 0.01   & 0.1     &  0.3    &  0.5  &  0.7  &   0.9    \\ \hline
$E_0$ (MeV)  & $-709$ & $-687$ & $-189$ & ~100~ & ~133~  & ~138~ &  ~139~ \\
$\Delta M$ (MeV)& $-848$ & $-826$ & $-328$ & $-39$ & $-6$ & $-1$  &  0 \\
 \hline
\end{tabular}
\end{center}
\end{table}

\begin{table}[ht]
\begin{center}
\caption{\label{result3}
%
Mass shifts computed for non-strange mesons with quantum numbers $IJ^-(I=0,1;J=0,1)$ using the transition matrix constructed from $T_{2}$ in Eq.\,\eqref{T2}: $f=0$, $\gamma=6.95$, $R_0=1$\,fm.  ($\eta$ is an isospin 0 partner to the pion; and all dimensioned quantities are listed in MeV.)}
\begin{tabular}{ccccc} \hline
states$(IJ^P)$ & $\pi(10^-)$ & $\rho(11^-)$ &  $\omega(01^-)$ &  $\eta(00^-)$ \\ \hline
bare mass (Theo.) & 139.0  &  772.7   &  701.9   & 669.5 \\ \hline
$\pi\pi$ & - & $-69.2$ & -  &  -  \\
$\pi\rho$ & $-318.2$ & - & $-231.2$ & - \\
$\pi\omega$ & -  & $-69.8$ &  -  &  -  \\
$\eta\rho$ & -  & $-52.3$ &  -  &  -  \\
$\rho\rho$ &  - & $-202.1$ &  -  &  $-267.8$  \\
$\rho\omega$ & $-280.1$  &  -  &  -  &  -  \\
$\eta\omega$ &  -  &  -  &  $-59.5$  &  - \\
$\omega\omega$ &  -   &  -  &  -  & $-91.3$ \\
$K \bar{K}$ &  -  &  $-22.1$  &  $-24.3$  &  -  \\
$K\bar{K}^{\star}(\bar{K}K^{\star})$ & $-114.0$ & $-38.2$  &  $-42.5$  & $-67.9$ \\
$K^{\star}\bar{K}^{\star}$ & $-215.2$  &  $-128.0$  &  $-147.4$  &  $-121.5$  \\ \hline
Total mass shift & ~$-927.5$~  & ~$-581.7$~  & ~$-504.9$~  & ~$-548.5$~  \\
 \hline
\end{tabular}
\end{center}
\end{table}

\subsubsection{Improvement Two}
Equally, the creation of quark-antiquark pairs should become less likely as the distance from the bare-hadron source is increased. This property is expressed in the following formula:
\begin{align} \label{T2}
T_2&= -3\gamma\sum_{m}\langle 1m1(-m)|00\rangle\int
d\mathbf{r_3}d\mathbf{r_4}(\frac{1}{2\pi})^{\frac{3}{2}}ir2^{-\frac{5}{2}}f^{-5}
\nonumber \\
 & Y_{1m}(\hat{\mathbf{r}})
 {\rm e}^{-\frac{\mathbf{r}^2}{4f^2}}
 {\rm e}^{-\frac{R_{AV}^2}{R_0^2}}\chi_{1-m}^{34}\phi_{0}^{34}
 \omega_{0}^{34}b_3^{\dagger}(\mathbf{r_3})d_4^{\dagger}(\mathbf{r_4}),
\end{align}
via the damping factor $e^{-R_{AV}^2/R_0^2}$.  Here, $R_{AV}$ is the relative distance between the source particle and quark-antiquark pair in the vacuum:
\begin{subequations}
\begin{align}
R_{AV}&= R_A-R_V;\\
R_A&=\frac{m_1\mathbf{r_1}+m_2\mathbf{r_2}}{m_1+m_2}; \\
R_V&= \frac{m_3\mathbf{r_3}+m_4\mathbf{r_4}}{m_3+m_4}=\frac{\mathbf{r_3}+\mathbf{r_4}}{2}(m_3=m_4).
\end{align}
\end{subequations}
A natural value for this production radius is $R_0$ $\approx$ 1\,fm, \emph{viz}.\ a typical hadronic size.  Table\,\ref{result3} shows the computed mass shifts in this situation: $f=0$, $\gamma=6.95$ and $R_0=1$\,fm.  The effect of the factor $e^{-R_{AV}^2/R_0^2}$ is to reduce the original mass-shifts  by roughly 50\%.

\subsubsection{Improvement Three: Combined Effect}
Based on the observations made above, both corrections to the $^{3}P_{0}$ pair-creation model should be considered
simultaneously when incorporating meson-loops.  Therefore, we build the complete $2q\to4q$ transition operator using Eq.\,\eqref{T2}, wherein now all three parameters, $\gamma$, $f$, $R_{0}$, are nonzero and active.  Table~\ref{all} shows the $f$-dependence of the eigen-energies and mass shifts obtained with $\gamma=6.95$, $R_0=1\,$fm.  Evidently, for each bound-state the mass-shift drops rapidly as $f$ increases; and, within our framework, the best value of $f$ can only be determined from data.

\begin{table*}[!t]
\begin{center}
\caption{$f$-dependence of meson masses and mass shifts, in MeV, obtained with the transition operator built from $T_2$ in Eq.\,\eqref{T2}, using $\gamma=6.95$, $R_0=1\,$fm.  Beginning with column\,2, each pair of columns reveals the channel, and the mass and mass-shift it yields as a function of $f$.
\label{all}}
\begin{tabular}{ccccccccc}
\hline
($\pi$) ~~$f$~~~~ &~~~~~~$\pi\rho$~~~~~  &~~~~~$\Delta M$~~~~~
&~~~~~$\rho\omega$~~~~~ &~~~~~~$\Delta M$~~~~~
&~~~~$KK^{\star}$ ~~~~~&~~~~$\Delta M$ ~~~~~&~~~~$K^{\star}K^{\star}$~~~~~ &~~~~~$\Delta M$~~~~~\\
 \hline
$0.1$   &46.4   &-92.6  &61.8 &-77.2 &119.2 &-19.8 &102.9 &-36.1
\\
$0.3$ &131.5 &-7.5 &132.9 &-6.1 &138.3 &-0.7 &137.7 &-1.3 \\
$0.5$ &138.1 &-0.9 &138.3 &-0.7 &138.9 &-0.1 &138.9 &-0.1 \\
$0.7$ &138.9 &-0.1 &138.9 &-0.1 &139.0 &0.0  &139.0 &0.0 \\
$0.9$ &139.0 &0.0  &139.0 &0.0  &139.0 &0.0  &139.0 &0.0 \\
 \hline
\end{tabular}\vspace*{2ex}

\begin{tabular}{ccccccccccccccc}\hline
($\rho$) ~~$f$~~~~ &~~$\pi\pi$~~  &~~$\Delta M$~~ &~~$\pi\omega$~~
&~~$\Delta M$~~ &~~$\eta\rho$~~ &~~$\Delta M$~~ &~~$\rho\rho$~~
&~~$\Delta M$~~ &~~$K K$~~&$\Delta M$~~
&~~$KK^{\star}$~~ &~~$\Delta M$~~&~~$K^{\star}K^{\star}$~~ &~~$\Delta M$~~\\
 \hline
$0.1$  &725.2 &-47.5 &735.4 &-37.3 &747.3 &-25.4 &678.7 &-94.0
&764.7 &-8.0 &759.5 &-13.2 &730.8 &-41.9 \\
$0.3$ &763.1 &-9.6 &766.7 &-6.0 &769.2 &-3.5 &759.5 &-13.2 &772.1
&-0.6 &771.8 &-0.9 &769.8 &-2.9\\
$0.5$ &771.8 &-0.9 &771.8 &-0.9 &772.2 &-0.5 &770.7 &-2.0 &772.7
&0.0 &772.6 &-0.1 &772.4 &-0.3\\
$0.7$&772.6 &-0.1 &772.6 &-0.1 &772.6 &-0.1 &772.4 &-0.3 &772.7
&0.0
&772.7 &0.0 &772.7 &0.0\\
$0.9$&772.7 &0.0 &772.7 &0.0 &772.7 &0.0 &772.7 &0.0 &772.7 &0.0
&772.7 &0.0
&772.7 &0.0\\
 \hline
\end{tabular}\vspace*{2ex}

\begin{tabular}{ccccccccccc}\hline
($\omega$) ~~$f$ ~~~~~~&$\pi\rho$~~~~~ &~~~~$\Delta M$~~~~
&~~~~$\eta\omega$~~~~ &~~~~$\Delta M$~~~~ &~~~~$KK$~~~~
&~~~~$\Delta M$~~~~ &~~~~$KK^{\star}$~~~~
 &~~~~$\Delta M$~~~~ &~~~~$K^{\star}K^{\star}$~~~~ &~~~~$\Delta M$~~~~\\
 \hline
$0.1$   &590.5 &-111.4 &674.7 &-27.2 &693.8  &-8.1  &688.4 &-13.5 &657.7 &-44.2\\
$0.3$   &685.9 &-16.0 &698.3  &-3.6 &701.3 &-0.6 &701.0    &-0.9 &699.1 &-2.8\\
$0.5$   &699.6 &-2.3 &701.4   &-0.5 &701.8 &-0.1 &701.8    &-0.1  &701.6 &-0.3\\
$0.7$   &701.5 &-0.4 &701.8   &-0.1  &701.9 &0.0  &701.9 &0.0  &701.8    &-0.1\\
$0.9$   &701.8 &-0.1 &701.9   &0.0   &701.9 &0.0  &701.9 &0.0  &701.9    &0.0\\
\hline
\end{tabular}\vspace*{2ex}

\begin{tabular}{ccccccccc}\hline
($\eta$) ~~$f$~~~~ &~~~~$\rho\rho$~~~~  &~~~~$\Delta M$~~~~
&~~~~$\omega\omega$~~~~ &~~~~$\Delta M$~~~~
&~~~~$KK^{\star}$~~~~ &~~~~$\Delta M$~~~~ &~~~~$K^{\star}K^{\star}$~~~~ &~~~~$\Delta M$~~~~\\
 \hline
$0.1$    &559.1 &-110.4 &629.7 &-39.8 &649.1 &-20.4 &634.6 &-34.9    \\
$0.3$    &655.4 &-14.1 &664.5 &-5.0  &668.2  &-1.3  &667.3 &-2.2  \\
$0.5$    &667.5  &-2.0  &668.8 &-0.7  &669.4 &-0.1 &669.3 &-0.2  \\
$0.7$    &669.1  &-0.4  &669.4 &-0.1 &669.5 &0.0 &669.5 &0.0\\
$0.9$    &669.4  &-0.1  &669.5 &0.0 &669.5 &0.0 &669.5 &0.0 \\
 \hline
\end{tabular}
\end{center}
\end{table*}

\begin{table}[htb]
\begin{center}
\renewcommand\arraystretch{1.0}
\caption{ \label{result4}
%
(A) -- Mass shifts computed for non-strange mesons with quantum numbers $IJ^-(I=0,1;J=0,1)$ using the transition matrix constructed from $T_{2}$ in Eq.\,\eqref{T2}: $f=0.5$, $\gamma=32.2$, $R_0=1$\,fm.
(B) -- Same as above, except that instead of the unquenched values in Table~\ref{modelparameters}, we used $\alpha_0=3.85$ (5\% increase) and $\Delta=-58.3\,$MeV (25\% increase).
($\eta$ is an isospin 0 partner to the pion; and all dimensioned quantities are listed in MeV.)}
\begin{tabular}{lccccc} \hline
(A) & \,$(IJ^P)$ & $\pi(10^-)$ & $\rho(11^-)$ &  $\omega(01^-)$ &  $\eta(00^-)$ \\ \hline
& bare mass (Theo.) & 139.0  &  772.7   &  701.9   & 669.5 \\ \hline
& $\pi\pi$ & - & $-18.0$ & -  &  -  \\
& $\pi\rho$ & $-18.3$ & - & $-45.5$ & - \\
& $\pi\omega$ & -  & $-19.1$ &  -  &  -  \\
& $\eta\rho$ & -  & $-11.4$ &  -  &  -  \\
& $\rho\rho$ &  - & $-41.7$ &  -  &  $-42.5$  \\
& $\rho\omega$ & $-15.4$  &  -  &  -  &  -  \\
& $\eta\omega$ &  -  &  -  &  $-10.9$  &  - \\
& $\omega\omega$ &  -   &  -  &  -  & $-15.3$ \\
& $K \bar{K}$ &  -  &  $-1.3$  &  $-1.2$  &  -  \\
& $K\bar{K}^{\star}(\bar{K}K^{\star})$ & $-1.3$ & $-2.0$  &  $-1.9$  & $2.7$ \\
& $K^{\star}\bar{K}^{\star}$ & $-2.4$  &  $-6.5$  &  $-6.0$  &  $-4.6$  \\
\hline
& Total mass shift & ~$-37.4$~  & ~$-100.0$~  & ~$-65.5$~  & ~$-65.1$~  \\
\hline
& unquenched mass &101.6 &672.7 &636.4 &604.4 \\
 \hline
\end{tabular}
\end{center}

\medskip

\begin{tabular}{lccccc} \hline
(B) & state\,$(IJ^P)$ & $\pi(10^-)$ & $\rho(11^-)$ &  $\omega(01^-)$ &  $\eta(00^-)$ \\ \hline
& bare mass (Theo.) & 172.7  &  869.5   &  798.5   & 747.8 \\ \hline
& $\pi\pi$ & - & $-21.4$ & -  &  -  \\
& $\pi\rho$ & $-16.3$ & - & $-42.3$ & - \\
& $\pi\omega$ & -  & $-17.2$ &  -  &  -  \\
& $\eta\rho$ & -  & $-10.6$ &  -  &  -  \\
& $\rho\rho$ &  - & $-38.9$ &  -  &  $-39.1$  \\
& $\rho\omega$ & $-13.8$  &  -  &  -  &  -  \\
& $\eta\omega$ &  -  &  -  &  $-10.2$  &  - \\
& $\omega\omega$ &  -   &  -  &  -  & $-13.9$ \\
& $K \bar{K}$ &  -  &  $-1.3$  &  $-1.2$  &  -  \\
& $K\bar{K}^{\star}(\bar{K}K^{\star})$ & $-1.2$ & $-2.3$  &  $-1.8$  & $-2.4$ \\
& $K^{\star}\bar{K}^{\star}$ & $-2.1$  &  $-6.1$  &  $-5.7$  &  $-4.3$  \\
\hline
& Total mass shift & ~$-33.4$~  & ~$-97.8$~  & ~$-61.2$~  & ~$-59.7$~  \\
\hline
& unquenched mass &139.3 &771.7 &737.3 &688.1 \\
 \hline
%
\end{tabular}

\end{table}

Given that the $\rho$-meson is properly regarded as primarily a $q\bar{q}$ meson, something we shall subsequently verify in our framework, and the $\rho\to \pi\pi$ branching fraction is 100\%, we fix $f$, $\gamma$ by fitting the decay width $\Gamma_{\rho \rightarrow \pi\pi} = 150\,$MeV and requiring that all mass shifts be reasonable, \emph{i.e}.\ neither too large, nor too small, and qualitatively consistent with field theory estimates \cite{Pichowsky:1999mu}.  In this way, we find:
\begin{equation}
\gamma = 32.2,\; \quad f=0.5\,\mbox{fm},
\end{equation}
with the associated meson mass shifts listed in Table\,\ref{result4}.A.   Plainly, our modified $ ^3P_0$ pair-creation model generates modest unquenching corrections, with mass renormalizations being just 10-25\% of a given meson's bare mass.

\begin{figure}[t]
\centerline{\includegraphics[width=0.4\textwidth]{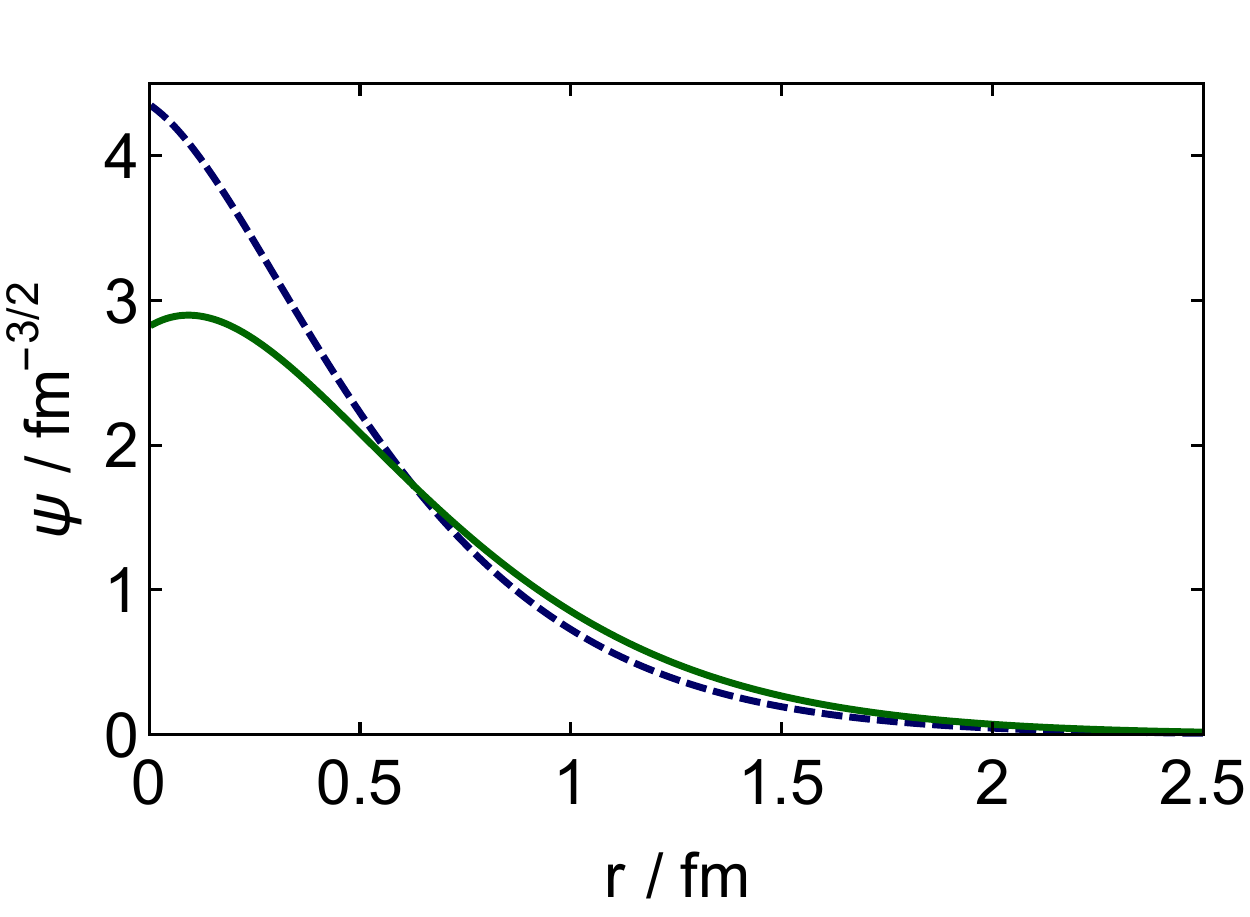}}
\caption{\label{F1}
Computed two-body quark-antiquark wave functions: solid (green) curve, $\rho$-meson; and dashed (blue) curve, $\omega$-meson.}
\end{figure}

We know that in QCD either the conservation of symmetries or the way they are broken is very important.  Applied to the case at hand, this means that certain patterns in the quark-antiquark and meson-meson couplings should be observed.  For instance, the $\rho$ and $\omega$ mesons differ only in their isospin ($I=1$ and $I=0$, respectively) and thus their couplings to $K\bar{K}$ and $K K^*$ should be identical.
Another example is the hidden channel $\rho \rightarrow \omega \pi$, in which the isospin coupling is $0\otimes1 \rightarrow 1$; and, complementing this, the isospin coupling for $\omega \rightarrow \rho \pi$ is $|00\rangle=1\otimes1=\frac{1}{\sqrt{3}}(|11\rangle|1-1\rangle-|10\rangle|10\rangle+|1-1\rangle|11\rangle)$.  Therefore, the $\omega \pi$ contribution to  the $\rho$-meson's mass shift should be equal to one-third of the $\rho \pi$ contribution to the $\omega$-meson's shift.
Finally, the $\omega \eta$ contribution to $\omega$ should equal the $\rho \eta$ contribution to $\rho$.
From Table \ref{result4}.A, one observes that the isospin-symmetry results are broadly respected.  However, there are some small discrepancies, which are dynamical in origin: as displayed in Fig.\,\ref{F1}, Goldstone boson exchange in the chiral quark model produces noticeable isospin-symmetry-breaking differences between the $\rho$ and $\omega$ wave functions.

It is worth highlighting here that the inability of the na\"{\i}ve chiral quark model to describe the $\rho-\omega$ splitting has long been known.  One proposal for solving the issue is inclusion of an explicit isospin-dependent mechanism in the light quark sector \cite{Vijande:2009pu}.  We have seen herein that the magnitude of the $\rho-\omega$ mass splitting can be reconciled with experiment when the contribution of meson-loops is included in a physically sound manner.  However, the level ordering remains incorrect.  As with much in the quark model treatment of light mesons, this devolves into an issue of fine tuning.
%
Notably, quantum field theory provides a different resolution \cite{Pichowsky:1999mu}, without fine-tuning, because it preserves the near isospin-symmetry of QCD.

In concluding this subsection, let us mention that a complete calculation that incorporates contributions from all possible multiple hadron intermediate states is beyond the scope of this work.  However, the improvements to the $^{3}P_{0}$ transition operator implemented herein ensure that the contributions of higher-mass intermediate states are small and hence the calculation should exhibit rapid convergence, making our results meaningful.

\subsection{Measured Masses and Four-Quark Components}
Notably, although the mass shifts reported in Table~\ref{result4}.A are sensible, they destroy agreement with the empirical masses.  This is because the parameters in Table~\ref{modelparameters} were determined by fitting the meson spectrum, without allowing room for $(q\bar q)^2$ components.  As a final exercise, therefore, we choose to \emph{illustrate} a remedy.  To that end, we adjust the OGE parameter $\alpha_0$ and confinement parameter $\Delta$ in order to increase the quenched masses of the $\pi$ and $\rho$ such that unquenching delivers the empirical masses, an outcome achieved with:
\begin{equation}
\alpha_0 = 3.85\,,\quad
\Delta = -58.3\,\mbox{MeV}.
\end{equation}
The results are listed in Table~\ref{result4}.B.  Evidently, the sizes of the mass shifts are not very sensitive to these parameters in the potential.  Having made our point, we leave for the future a complete refit of the parameters in Table~\ref{modelparameters} in order to arrive finally at a fully unquenched quark model.

Having produced the results in Table~\ref{result4}.B, it is meaningful to compute the strength of all $(q\bar q)^2$ contributions to each unquenched quark model state.  Our results are listed in Table~\ref{components}: with more intermediate states available, and a sizeable coupling to the $\pi\pi$ channel, the $\rho$-meson possesses the largest $(q\bar q)^2$ component.


\section{Summary}
\label{epilogue}
A coupled-channels calculation of the spectrum of light mesons with quantum numbers $IJ^{P=-}$, $(I,J)=0,1$, has
been presented.  Within a chiral quark model, the $q\bar q$ and $(q\bar q)^2$ masses and wave functions were obtained by solving the Schr\"odinger equation using the Gaussian Expansion Method.  The coupling between two- and four-quark configurations was realized through a modified version of the transition operator in the $^{3}P_{0}$ decay model.  This new version allows us to recover the original in a particular limit and compare the mass-shifts generated by unquenching in a variety of scenarios.

Solving the coupled-channels problem using the original $^{3}P_{0}$ operator, we found that the mass shifts for the $\pi, \rho, \omega, \eta$ mesons are very large and negative, an outcome which seriously undermines the quenched model.  Such a conclusion regarding the validity of that model is unexpected because it provides a reasonable description of many hadrons and their decays.  We judged, therefore, that the simple $^{3}P_{0}$ transition operator needed modification so as to ensure that hadron-loop effects do not generate mass shifts that exceed roughly 10-20\% of the hadron bare masses computed in the chiral quark model.

We incorporated two simple, physically motivated improvements into the $^3P_0$ transition operator, ensuring that: (\emph{i}) intermediate dressing-states with large momentum are suppressed; and (\emph{ii}) quark-antiquark creation near the hadron source is favored.  With these improvements, the mass shift in each channel considered is reduced by an order-of-magnitude or more, so that the corrected results amount to a shift of only 10-20\% of the quenched mass value.  These improvements ensure additionally that high-mass intermediate states are damped, according to their mass, and hence that the sum of meson-loop corrections converges quickly, as it typically does in realistic quantum field theory calculations.

It is also worth mentioning both that our modified operator fulfills some transition coupling rules, which are imposed by isospin symmetry; and, by illustration, we showed that the parameters of the na\"{\i}ve chiral quark model may be adjusted so that a quantitatively useful unquenched version can be developed in future.


\begin{table}[!t]
\renewcommand\arraystretch{1.0}
\caption{\label{components}
Fractions (\%) of two- and four-quark components in the unquenched mesons, computed using the framework developed for Table~\ref{result4}.B.
%
}
\begin{tabular}{l|cccc}\hline
                        & $\pi$ & $\rho$ & $\omega$ & $\eta$ \\\hline
bare $q \bar q$ & 97.8 & 74.3 & 92.7 & 95.3 \\\hline
$\pi\pi$           &  -     & 18.4 &   -    &  - \\
$\pi \rho$      & 1.2    &    -     & 5.9 & - \\
$\pi \omega$    &   -    & 3.0   &   -    &- \\
$\eta \rho$     &    -     & 0.8   &   -    & -\\
$\eta \omega$  &  -     &  -      &  0.8  &- \\
$\rho\rho$  &      -      & 2.9   &    -    & 3.1 \\
$\rho\omega$   & 0.8 &   -     &    -    & -\\
$\omega\omega$  & -  &  -      &  -      & 1.1 \\
$KK$    &            -       & 0.1  & 0.1  &  -\\
$K K^\ast$  &     0.1   & 0.2  & 0.1  & 0.2 \\
$K^\ast K^\ast$ & 0.1 & 0.3 & 0.4  & 0.3   \\\hline
\end{tabular}
\end{table}


\acknowledgments
We are grateful for insightful comments from C.~Chen, Y.~Lu, C.~Shi and S.-S. Xu.
Work supported by:
National Natural Science Foundation of China, under Grant Nos.\ 11535005, 11775118, and 11205091;
U.S.\ Department of Energy, Office of Science, Office of Nuclear Physics, under contract no.~DE-AC02-06CH11357;
Chinese Ministry of Education, under the \emph{International Distinguished Professor} programme;
European Union's Horizon 2020 research and innovation programme under the Marie Sk\l{}odowska-Curie Grant Agreement No.\ 665919;
Spanish MINECO's Juan de la Cierva-Incorporaci\'on programme, Grant Agreement No. IJCI-2016-30028;
and Spanish Ministerio de Econom\'ia, Industria y Competitividad under Contract Nos.\ FPA2014-55613-P and SEV-2016-0588.



\begin{thebibliography}{10}

\bibitem{Eichten:1978tg}
E.~Eichten, K.~Gottfried, T.~Kinoshita, K.~D. Lane and T.-M. Yan,
\newblock Phys. Rev. D {\bf 17}, 3090 (1978),
\newblock [Erratum: Phys. Rev. D \textbf{21}, 313 (1980)].

\bibitem{Eichten:1979ms}
E.~Eichten, K.~Gottfried, T.~Kinoshita, K.~D. Lane and T.-M. Yan,
\newblock Phys. Rev. D {\bf 21}, 203 (1980).

\bibitem{Gupta:1982kp}
S.~N. Gupta, S.~F. Radford and W.~W. Repko,
\newblock Phys. Rev. D {\bf 26}, 3305 (1982).

\bibitem{Gupta:1983we}
S.~N. Gupta, S.~F. Radford and W.~W. Repko,
\newblock Phys. Rev. D {\bf 28}, 1716 (1983).

\bibitem{Gupta:1984jb}
S.~N. Gupta, S.~F. Radford and W.~W. Repko,
\newblock Phys. Rev. D {\bf 30}, 2424 (1984).

\bibitem{Gupta:1984um}
S.~N. Gupta, S.~F. Radford and W.~W. Repko,
\newblock Phys. Rev. D {\bf 31}, 160 (1985).

\bibitem{Kwong:1987ak}
W.~Kwong, P.~B. Mackenzie, R.~Rosenfeld and J.~L. Rosner,
\newblock Phys. Rev. D {\bf 37}, 3210 (1988).

\bibitem{Kwong:1988ae}
W.~Kwong and J.~L. Rosner,
\newblock Phys. Rev. D {\bf 38}, 279 (1988).

\bibitem{Barnes:1996ff}
T.~Barnes, F.~E. Close, P.~R. Page and E.~S. Swanson,
\newblock Phys. Rev. D {\bf 55}, 4157 (1997).

\bibitem{Ebert:2002pp}
D.~Ebert, R.~N. Faustov and V.~O. Galkin,
\newblock Phys. Rev. D {\bf 67}, 014027 (2003).

\bibitem{Radford:2007vd}
S.~F. Radford and W.~W. Repko,
\newblock Phys. Rev. D {\bf 75}, 074031 (2007).

\bibitem{Eichten:2007qx}
E.~Eichten, S.~Godfrey, H.~Mahlke and J.~L. Rosner,
\newblock Rev. Mod. Phys. {\bf 80}, 1161 (2008).

\bibitem{Segovia:2008zz}
J.~Segovia, A.~M. Yasser, D.~R. Entem and F.~Fern{\'a}ndez,
\newblock Phys. Rev. D {\bf 78}, 114033 (2008).

\bibitem{Danilkin:2009hr}
I.~V. Danilkin and {\relax Yu}.~A. Simonov,
\newblock Phys. Rev. D {\bf 81}, 074027 (2010).

\bibitem{Ferretti:2013vua}
J.~Ferretti and E.~Santopinto,
\newblock Phys. Rev. D {\bf 90}, 094022 (2014).

\bibitem{Segovia:2013wma}
J.~Segovia, D.~R. Entem, F.~Fern{\'a}ndez and E.~Hernandez,
\newblock Int. J. Mod. Phys. E {\bf 22}, 1330026 (2013).

\bibitem{Godfrey:2015dia}
S.~Godfrey and K.~Moats,
\newblock Phys. Rev. D {\bf 92}, 054034 (2015).

\bibitem{Segovia:2016xqb}
J.~Segovia, P.~G. Ortega, D.~R. Entem and F.~Fern{\'a}ndez,
\newblock Phys. Rev. D {\bf 93}, 074027 (2016).

\bibitem{Godfrey:1985xj}
S.~Godfrey and N.~Isgur,
\newblock Phys. Rev. D {\bf 32}, 189 (1985).

\bibitem{Vijande:2004he}
J.~Vijande, F.~Fern{\'a}ndez and A.~Valcarce,
\newblock J. Phys. G {\bf 31}, 481 (2005).

\bibitem{Segovia:2008zza}
J.~Segovia, D.~R. Entem and F.~Fern{\'a}ndez,
\newblock Phys. Lett. B {\bf 662}, 33 (2008).

\bibitem{Choi:2003ue}
S.~K. Choi {\em et~al.},
\newblock Phys. Rev. Lett. {\bf 91}, 262001 (2003).

\bibitem{Brambilla:2010cs}
N.~Brambilla {\em et~al.},
\newblock Eur. Phys. J. C {\bf 71}, 1534 (2011).

\bibitem{Belle:2011aa}
A.~Bondar {\em et~al.},
\newblock Phys. Rev. Lett. {\bf 108}, 122001 (2012).

\bibitem{Ablikim:2013mio}
M.~Ablikim {\em et~al.},
\newblock Phys. Rev. Lett. {\bf 110}, 252001 (2013).

\bibitem{Tornqvist:1979hx}
N.~A. Tornqvist,
\newblock Annals Phys. {\bf 123}, 1 (1979).

\bibitem{Ono:1983rd}
S.~Ono and N.~A. Tornqvist,
\newblock Z. Phys. C {\bf 23}, 59 (1984).

\bibitem{Tornqvist:1984xz}
N.~A. Tornqvist,
\newblock Acta Phys. Polon. B {\bf 16}, 503 (1985),
\newblock [Erratum: Acta Phys. Polon. \textbf{B} 16, 683 (1985)].

\bibitem{Ping:2012zz}
J.~Ping, C.~Deng, H.~Huang, F.-F. Dong and F.~Wang,
\newblock EPJ Web Conf. {\bf 20}, 01007 (2012).

\bibitem{Ortega:2016mms}
P.~G. Ortega, J.~Segovia, D.~R. Entem and F.~Fern{\'a}ndez,
\newblock Phys. Rev. D {\bf 94}, 074037 (2016).

\bibitem{Ortega:2016pgg}
P.~G. Ortega, J.~Segovia, D.~R. Entem and F.~Fern{\'a}ndez,
\newblock Phys. Rev. D {\bf 95}, 034010 (2017).

\bibitem{Ortega:2016hde}
P.~G. Ortega, J.~Segovia, D.~R. Entem and F.~Fern{\'a}ndez,
\newblock Phys. Rev. D {\bf 94}, 114018 (2016).

\bibitem{Ortega:2017qmg}
P.~G. Ortega, J.~Segovia, D.~R. Entem and F.~Fern{\'a}ndez,
\newblock (2017),
\newblock {\emph{Charmonium resonances in the 3.9 GeV/$c^2$ energy region and
  the $X(3915)/X(3930)$ puzzle}, arXiv:1706.02639 [hep-ph]},.

\bibitem{Geiger:1989yc}
P.~Geiger and N.~Isgur,
\newblock Phys. Rev. D {\bf 41}, 1595 (1990).

\bibitem{Micu:1968mk}
L.~Micu,
\newblock Nucl. Phys. B {\bf 10}, 521 (1969).

\bibitem{LeYaouanc:1972vsx}
A.~Le~Yaouanc, L.~Oliver, O.~P{\`e}ne and J.~C. Raynal,
\newblock Phys. Rev. D {\bf 8}, 2223 (1973).

\bibitem{LeYaouanc:1973ldf}
A.~Le~Yaouanc, L.~Oliver, O.~P{\`e}ne and J.-C. Raynal,
\newblock Phys. Rev. D {\bf 9}, 1415 (1974).

\bibitem{Barnes:2007xu}
T.~Barnes and E.~S. Swanson,
\newblock Phys. Rev. C {\bf 77}, 055206 (2008).

\bibitem{Ortega:2009hj}
P.~G. Ortega, J.~Segovia, D.~R. Entem and F.~Fern{\'a}ndez,
\newblock Phys. Rev. D {\bf 81}, 054023 (2010).

\bibitem{Li:2009ad}
B.-Q. Li, C.~Meng and K.-T. Chao,
\newblock Phys. Rev. D {\bf 80}, 014012 (2009).

\bibitem{Aubert:2008ba}
B.~Aubert {\em et~al.},
\newblock Phys. Rev. Lett. {\bf 101}, 071801 (2008), [0807.1086].

\bibitem{delAmoSanchez:2010kz}
P.~del Amo~Sanchez {\em et~al.},
\newblock Phys. Rev. {\bf D82}, 111102 (2010).

\bibitem{Lees:2011zp}
J.~P. Lees {\em et~al.},
\newblock Phys. Rev. D {\bf 84}, 091101 (2011).

\bibitem{Adachi:2011ji}
I.~Adachi {\em et~al.},
\newblock Phys. Rev. Lett. {\bf 108}, 032001 (2012).

\bibitem{Liu:2011yp}
J.-F. Liu and G.-J. Ding,
\newblock Eur. Phys. J. C {\bf 72}, 1981 (2012).

\bibitem{Geiger:1991qe}
P.~Geiger and N.~Isgur,
\newblock Phys. Rev. Lett. {\bf 67}, 1066 (1991).

\bibitem{Blundell:1995ev}
H.~G. Blundell and S.~Godfrey,
\newblock Phys. Rev. D {\bf 53}, 3700 (1996).

\bibitem{Hiyama:2003cu}
E.~Hiyama, Y.~Kino and M.~Kamimura,
\newblock Prog. Part. Nucl. Phys. {\bf 51}, 223 (2003).

\bibitem{Valcarce:2005em}
A.~Valcarce, H.~Garcilazo, F.~Fern{\'a}ndez and P.~Gonzalez,
\newblock Rept. Prog. Phys. {\bf 68}, 965 (2005).

\bibitem{Capstick:1986bm}
S.~Capstick and N.~Isgur,
\newblock Phys. Rev. D {\bf 34}, 2809 (1986).

\bibitem{Roberts:1992js}
W.~Roberts and B.~Silvestre-Brac,
\newblock Acta Phys. Austriaca {\bf 11}, 171 (1992).

\bibitem{Capstick:1993kb}
S.~Capstick and W.~Roberts,
\newblock Phys. Rev. D {\bf 49}, 4570 (1994).

\bibitem{Page:1995rh}
P.~R. Page,
\newblock Nucl. Phys. B {\bf 446}, 189 (1995).

\bibitem{Ackleh:1996yt}
E.~S. Ackleh, T.~Barnes and E.~S. Swanson,
\newblock Phys. Rev. D {\bf 54}, 6811 (1996).

\bibitem{Segovia:2012cd}
J.~Segovia, D.~R. Entem and F.~Fern{\'a}ndez,
\newblock Phys. Lett. B {\bf 715}, 322 (2012).

\bibitem{Jacob:1959at}
M.~Jacob and G.~C. Wick,
\newblock Annals Phys. {\bf 7}, 404 (1959),
\newblock [Annals Phys. \textbf{281}, 774 (2000)].

\bibitem{LeYaouanc:1977gm}
A.~Le~Yaouanc, L.~Oliver, O.~Pene and J.~C. Raynal,
\newblock Phys. Lett. B {\bf 72}, 57 (1977).

\bibitem{Pichowsky:1999mu}
M.~A. Pichowsky, S.~Walawalkar and S.~Capstick,
\newblock Phys. Rev. D {\bf 60}, 054030 (1999).

\bibitem{Vijande:2009pu}
J.~Vijande and A.~Valcarce,
\newblock Phys. Lett. B {\bf 677}, 36 (2009), [0905.1817].

\end{thebibliography}

\end{document}